\documentclass[12pt,preprint]{aastex}

\slugcomment{}

\shorttitle{3-D structure and distance of NGC 6781}
\shortauthors{Schwarz and Monteiro}

\begin{document}
\title{3-D Photoionization Structure and Distances of 
Planetary Nebulae III. NGC 6781}

\author{Hugo E. Schwarz$^1$, Hektor Monteiro$^{1,2}$}
\affil{1 Cerro Tololo Inter-American Observatory\altaffilmark{1}, 
Casilla 603, Colina El Pino S/N, La Serena, Chile}

\affil{2 Department of Physics and Astronomy, Georgia State University,  
1, Park Place South, Atlanta, GA 30302, USA}

\altaffiltext{1}{Cerro Tololo Inter-American Observatory, National 
Optical Astronomy Observatory, operated by the Association of Universities for
Research in Astronomy, Inc., under a cooperative agreement with the
National Science Foundation.}

\begin{abstract}

Continuing our series of papers on the three-dimensional (3-D)
structures of and accurate distances to Planetary Nebulae (PNe), we
present our study of the planetary nebula NGC\,6781. For this object
we construct a 3-D photoionization model and, using the constraints
provided by observational data from the literature we determine the
detailed 3-D structure of the nebula, the physical parameters of the
ionizing source and the first precise distance. The procedure consists
in simultaneously fitting all the observed emission line morphologies,
integrated intensities and the 2-D density map from the [SII] line
ratios to the parameters generated by the model, and in an iterative
way obtain the best fit for the central star parameters and the
distance to NGC\,6781, obtaining values of 950\,$\pm$\,143\,pc and
385\,L$_{\odot}$ for the distance and luminosity of the central star
respectively. Using theoretical evolutionary tracks of intermediate
and low mass stars, we derive the mass of the central star of
NGC\,6781 and its progenitor to be $0.60\,\pm0.03\,M_{\odot}$ and
$1.5\,\pm0.5\,M_{\odot}$ respectively.

\end{abstract}

\keywords{Planetary Nebula -- Interstellar medium -- photoionization 
modeling}

\section{Introduction}

Planetary nebulae are the end products of the evolution of stars with
masses below about $8 M_{\odot}$ (\cite{P84} and \cite{IR83}). The
importance of these objects extends beyond the understanding of how
the outer layers of stars end up forming the many observed
morphologies. Indeed, PNe have been used for many purposes, from
understanding basic atomic and plasma processes (\cite{A87}) to
determining chemical evolution of our Galaxy (\cite{MC03}). Recently
PNe are also being used to study galaxies other than our own,
providing powerful tools to determine distances, kinematics and
chemical properties of external galaxies (\cite{C03}) and even to
trace inter-cluster material as in \cite{FCJD04}. PNe have the
advantage that they are observable out to large distances due to their
luminous narrow emission lines, especially [OIII]500.7nm, being
useful also as standard candles through the use of the PN luminosity
function (for extensive discussion see \cite{JCF99}.

One of the most important problems in observational Galactic PNe
research is the difficulty in determining their distances and
three-dimensional structures. Observations always produce a 2-D
projection of their 3-D structure, and recovering the original
structure is not trivial. This is also made worse by the fact that
only crude distances, usually obtained from statistical methods on
large samples, can be determined.  Large uncertainties are generated
by the need to assume constancy of one parameter such as the nebular
size, (ionized) mass, flux etc. so that typical errors in the
distances to individual objects are of the order of a factor of 3 or
more. Very few nebulae have had individual accurate distances
determined.

Historically, PNe have been studied with empirical methods and one
dimensional photoionization models, leading to the above mentioned
problems. In contrast to this, our technique developed and described in
\cite{MSGH04} provides precise, self-consistently determined
distances, as well as the physical parameters for the central star and
gaseous nebula, for objects with sufficient observational
constraints. These objects can provide valuable calibration for
existing distance scales as well as self-consistently determined
physical and chemical quantities. For a detailed description of our
novel method see our previously published papers in this series and
especially the extensive explanatory appendix in the second paper 
\cite{MSGGH05}.

\begin{figure}[ht]
\includegraphics[width=\columnwidth]{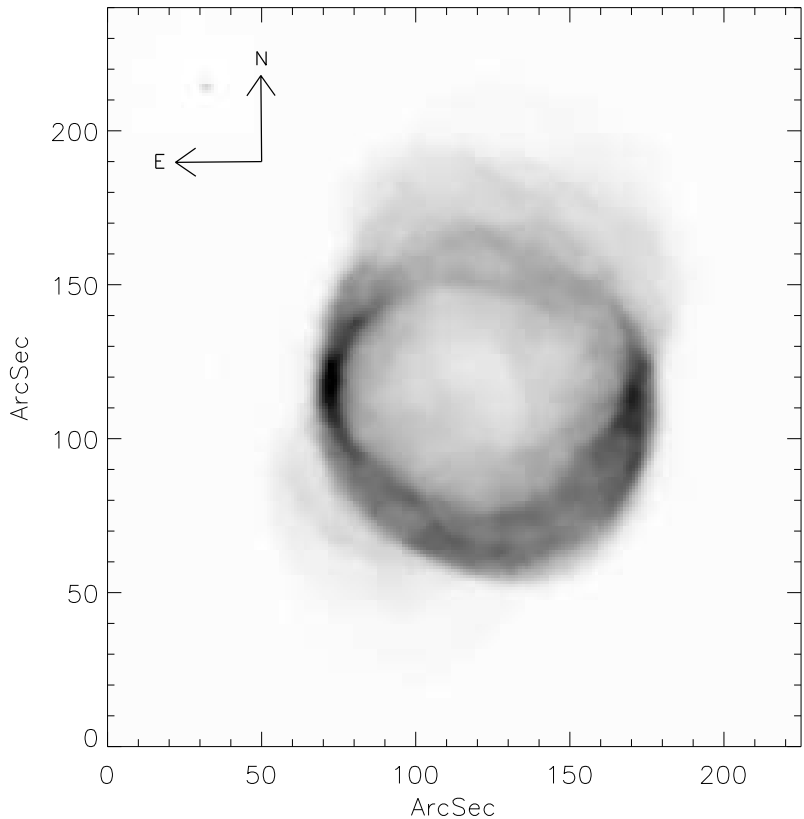}
\caption{Narrowband image of NGC\,6781 in the light of the [NII]658.4nm 
line. Image produced from the original of \cite{MPPS01}. \label{imagen6781}}
\end{figure}

In this work we focus on the PN NGC\,6781 (RA 19$^h$ 18$^m$ 28$^s$
Dec. +06$^o$ 32' 19'' 2000.) shown in Figure\,\ref{imagen6781}. This
is a PN whose main structure is observed as a 130'' diameter bright
shell of low ellipticity, double in parts and with fainter lobes
emanating at the N-S ends where the ring is double and fainter.  The
ring is brighter in the E-W directions as also shown in Figure\,1 of
\cite{MPPS01}. Their image is available as a fits file on the web at
http://www.ing.iac.es/~rcorradi and our Figure\,\ref{imagen6781} has been produced
using this image.

\cite{MPPS01} claim a possible faint halo extending out to about 3' by 4' 
surrounding NGC\,6781. Note that \cite{CSSP03} list the object in
their paper on the search for faint haloes around bright PNe, but do
not make any statement about having detected a halo according to their
criteria, which include the candidate halo having to be limb
brightened and\/or detached. The halo of \cite{MPPS01} is not limb
brightened and more likely to be the result of scattered light in the
instrument.

For a more detailed discussion of the morphology of NGC\,6781 based on
a set of narrow band images, as well as density, temperature, and
extinction maps see \cite{MPPS01}. 

An estimate for the luminosity, temperature, and mass of the CS of
NGC\,6781 based on statistical analysis is 127\,L$_{\odot}$, 105\,kK,
and 0.6\,M$_{\odot}$, respectively, taken from \cite{SVMG02}. The
average distance of the 12 literature values we found in \cite{AO92}
is 946\,pc with the individual values ranging from 500 to 1600\,pc, a
factor of more than three.

\begin{figure}[ht]

\includegraphics[width=\columnwidth]{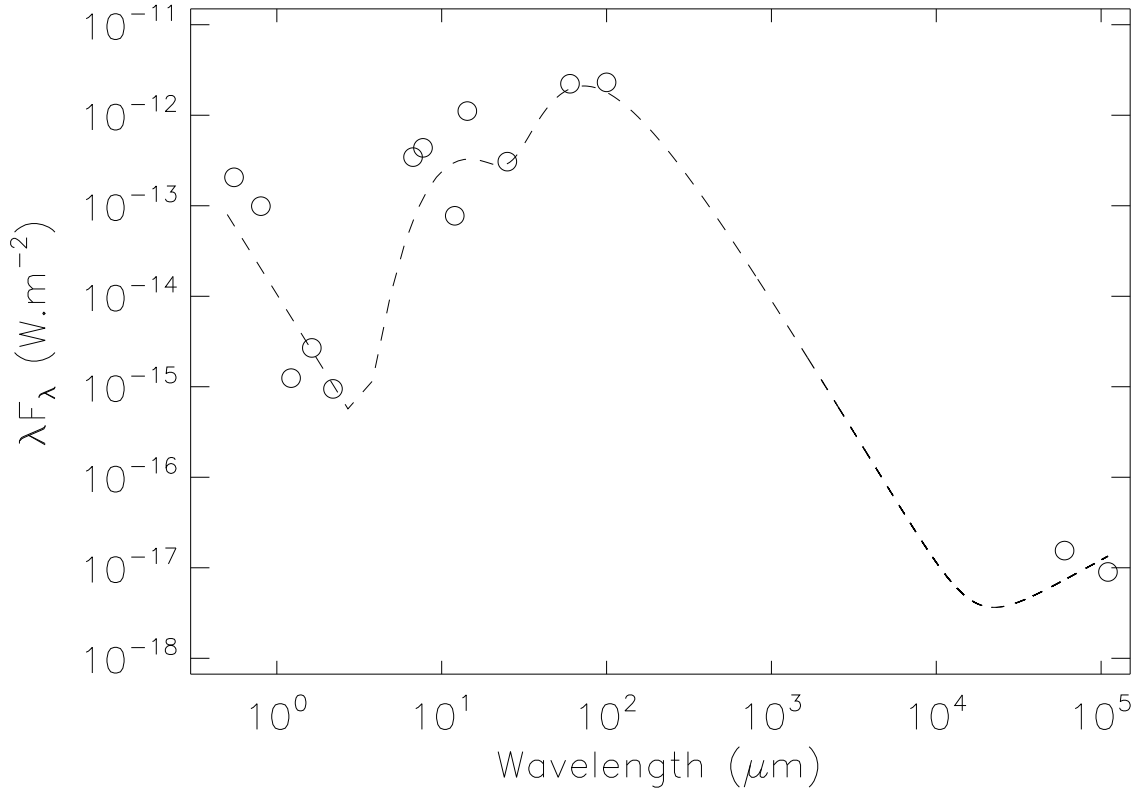}

\caption{
The graph shows the Spectral Energy Distribution (SED) of NGC\,6781 over more 
than five decades of wavelength interval. All values were obtained from 
the literature. The dotted curve is a combined blackbody and continuum 
radiation fit to the data. \label{sed6781}}
\end{figure} 

The Spectral Energy Distribution (SED) $\lambda$F($\lambda$) of
NGC\,6781 shows a broad blackbody-like spectrum between about 1 and
100\,$\mu$m, a radio tail out to 12\,cm, and a rise toward the blue
which probably comes from the hot central star.  Integrating the F($\lambda$) curve we
obtain a luminosity of L = 166\,d$^2$\,(kpc) L$_{\odot}$. The
luminosity of NGC\,6781 is therefore 150\,L$_{\odot}$ at its distance
of 950\,pc. Applying the usual correction according to \cite{MY87} we
obtain L\,=\,225\L$_{\odot}$.

Here we present our own modeling results using observational
data published in the literature as constraints and derive the 3-D
structure, chemical abundances, CS properties, and distance for
NGC\,6781 in a self-consistent manner.

In \S2 we discuss the observational data used as constraints for the
models. In \S3 we present the model results generated by the 3-D
photoionization code, and we discuss the derived quantities. In \S4
we give our overall conclusions and discuss possible discrepancies
with other determinations of parameters for this nebula.

\section{Observations}

The observational data that we used to constrain our model for
NGC\,6781 were taken from the literature.  \cite{MPPS01} presents narrow
band imaging of the nebula in the most important lines such as
H$\alpha$ ,H$\beta$, [OIII] and others. He also provides [SII] narrow
band images to obtain a spatially resolved density map, as well as the
H$\alpha$\/H$\beta$ extinction map of the nebula. We used their 2-D
density map to infer the 3-D density structure to be used as an input
for the photoionization model.

\cite{LLLB04} present deep optical spectra of medium resolution for
NGC 6781 and 11 other planetary nebulae. The observations were carried
out with a long-slit spectrograph covering from 360 to 800\,nm and
include all important emission lines.  Their observations are
particularly interesting because the objects were scanned with the
long-slit across the nebular surface by driving the telescope
differentially in right ascension. These observations then yield
average spectra for the whole nebula, which are in principle more
precise than single slit observations, and allow us to produce
emission line maps of the object. For details of the observations and
their reduction procedure as well as the full tables of line fluxes,
see \cite{LLLB04}. The most important lines used as constraints for
the model are listed in our Table\,\ref{mod-res} together with the
corresponding values from our model.

We also used the H$\alpha$\,+\,[NII] image from \cite{MPPS01} to
determine the size of NGC\,6781 thus using it as one of the
constraints for the distance obtained in our model calculations.

\section{Photoionization Models for NGC\,6781}

The photoionization code we used for the study of NGC\,6781 is the
Mocassin code described in full detail in \cite{EBSL03}. This
code allows for the same possibilities as the code used previously in
\cite{MMGV00} but is more sophisticated in that the diffuse radiation is
fully taken into account in an efficient manner. The previous code
also had this ability but the associated increase in computational
time was prohibitive.

The basic procedure adopted to study NGC\,6781 is the same,
independent of the code and has been fully described in the two
previous papers in this series (\cite{MSGH04} and (especially the
appendix of) \cite{MSGGH05}). In short, we gather as much
observational material as possible and use it to constrain our model
with many data simultaneously. Of particular interest are the total
line fluxes and line images corrected for reddening, and the line
diagnostic ratio maps. The structure adopted for the nebulae is
defined based on the observed density map, when available, or density
profiles from single slit observations plus the observed projected
morphology in several emission lines.

\begin{figure}[ht]
\includegraphics[width=\columnwidth]{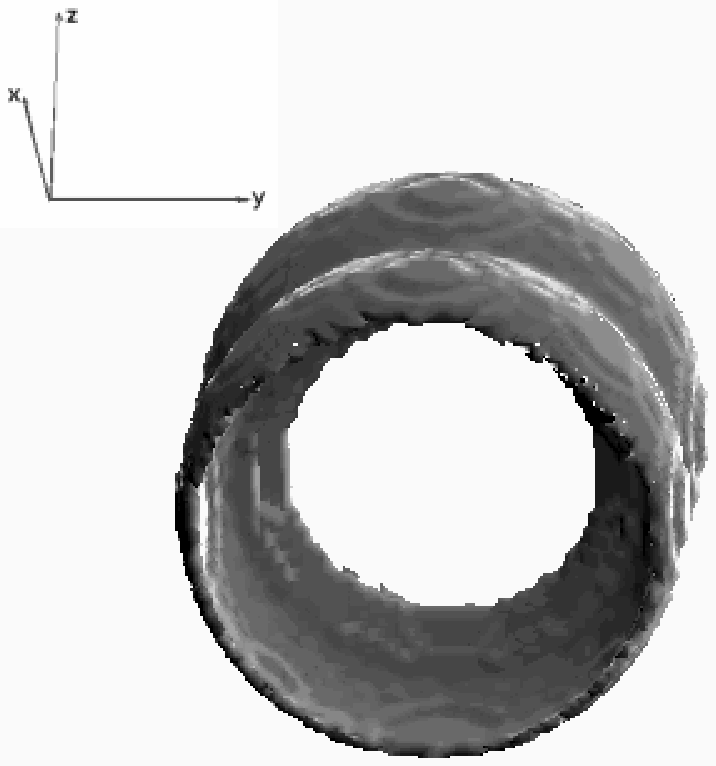}
\caption{The image shows an isodensity surface of the structure 
adopted for NGC\,6781 in its final fitted spatial orientation 
indicated in the upper left hand corner. \label{strut6781}}
\end{figure}

In the case of NGC\,6781 in particular, the initial structure was
based on density maps published by \cite{MPPS01}. It is clear from the
images and density map (from the [SII] doublet ratio) presented here,
that the density is lower in the central region than in the
main bright ring. This indicates that the structure must have lower
density material in the line of sight of those regions and therefore
the best structure to reproduce the observed projected morphology is
an open ended structure or hour-glass shape, which we therefore adopt
for this object. It is also clear from the images in many narrow band
filters that the material is highly clumpy, so to reproduce this we
include random density fluctuations in the adopted structure. The
final adopted structure in its best fitting orientation on the sky is
presented in Figure\,\ref{strut6781}.

\section{Model Results}

We present here the main results obtained from the photoionization
model constrained by the observational data. The integrated fluxes for
12 emission lines are given in Table\,\ref{mod-res}, as well as the
fitted abundances and ionizing star parameters.

The model fitting procedure which uses the model image size fitted to
the observed one for the line {[}NII{]}658.4nm, as well as the
absolute $H\beta$ flux, and the integrated fluxes of all other lines
gives a distance of 950$\pm$145\,pc for NGC\,6781. The error on this
distance has been computed in the same way as in our previous papers
in this series.
  
\begin{table}[ht]
\begin{center}
\caption{Observed and model line fluxes and model central star 
parameters for NGC\,6781.\label{mod-res}}
\begin{tabular}{lccc}\hline  &   {Observed}  &   {Model} & {Rel. Error} \\
\hline\hline 
\\
{$T_*$ (K)} & {-} & {123kK}                         &  {-}  \\
{$L_*/L_{\odot}$} & {-} & {385}                     &  {-}  \\
{log\,($g_*$)} & {-} & {7.0}                        &  {-}  \\
{Distance (pc)} & 500\,-\,1600 & 950            &   {$\pm$0.143} \\
{Density}   & {100-1400} & {100-1400}               &  {-}  \\
{He/H}      & {-}     &   {$1.25\times 10^{-1}$}    &  {-}  \\
{C/H}       & {-}     &   {$5.95\times 10^{-4}$}    &  {-}  \\
{N/H}       & {-}     &   {$9.75\times 10^{-5}$}    &  {-}  \\
{O/H}       & {-}     &   {$3.50\times 10^{-4}$}    &  {-}  \\
{Ne/H}      & {-}     &   {$7.10\times 10^{-5}$}    &  {-}  \\
{Ar/H}      & {-}     &   {$2.26\times 10^{-6}$}    &  {-}  \\
{S/H}       & {-}     &   {$0.28\times 10^{-5}$}    &  {-}  \\
{log($H\beta$)}       &   {-9.80}    &   {-9.84} & {-0.09}  \\
{[NeIII]\,386.8$^{a}$}&   {1.09}     &   {1.06} &  {-0.03}  \\
{[OIII]\,436.3}       &   {0.05}     &   {0.07} &  {0.28}   \\
{HeII\,468.6}         &   {0.09}     &   {0.09} &  {0.02}   \\
{[OIII]\,500.7}       &   {8.23}     &   {7.10} &  {-0.14}  \\
{[NII]\,575.5}        &   {0.07}     &   {0.07} &  {0.07}   \\
{HeI\,587.6}          &   {0.16}     &   {0.17} &  {0.09}   \\
{[OI]\,630.2}         &   {0.33}     &   {0.38} &  {0.15}   \\
{H$\alpha$656.3}      &   {2.65}     &   {2.86} &  {0.08}   \\
{[NII]\,658.4}        &   {3.96}     &   {3.88} &  {-0.02}  \\
{[SII]\,671.7}        &   {0.25}     &   {0.25} &  {-0.01}  \\
{[SII]\,673.1}        &   {0.21}     &   {0.21} &  {0.00}   \\     

\hline

\end{tabular}
\end{center}
a) Value obtained by \cite{KSB97}
\end{table}

\begin{figure*}[ht]

\begin{center}
\includegraphics[scale=0.75]{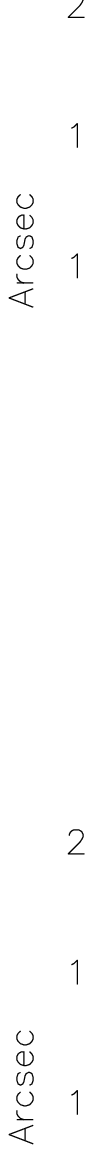}

\caption{
Images obtained from the projection at the observed angle of the data 
cubes of emissivities computed by the photoionization code for the most 
important emission lines. Compare these model images with the observed 
images in \cite{MPPS01}. \label{modims6781} }
\end{center}
\end{figure*}

In Fig. \ref{modims6781} we show the projected images obtained from
the fitted model. Notice that all major morphological features of the
object are well reproduced as well as the general ionization
stratification in the different emission lines. In particular the
images for [OIII]\,500.7 and HeII\,468.6 show very good agreement with
those obtained by \cite{MPPS01} as well as the more common
[NII]\,658.4\,nm and $H\alpha$.

\begin{figure}[ht]
\includegraphics[width=\columnwidth]{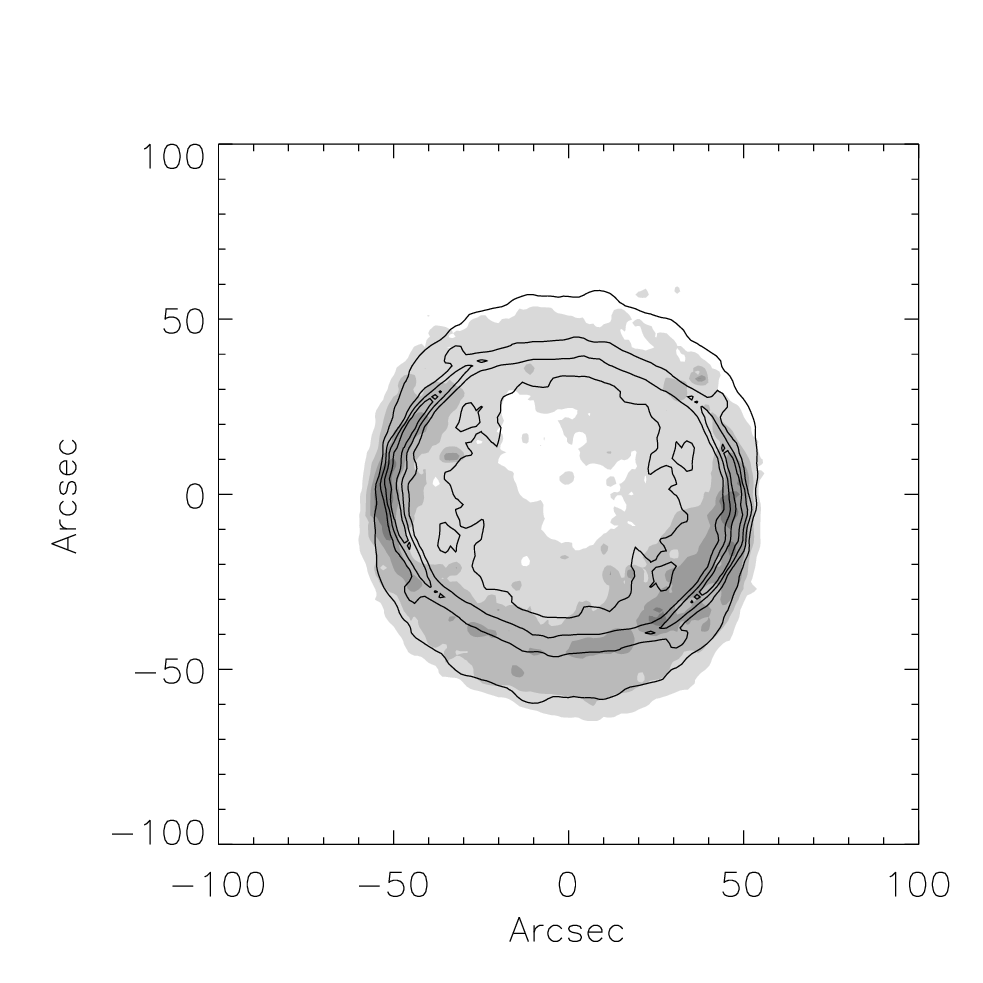}
\caption{
Image comparing the observed [NII]658.4\,nm narrow band image with the
contours of the equivalent image of the fitted model. Note the
similarity between the observed and modeled images. \label{n2comp6781}
}
\end{figure}

Figure\,\ref{n2comp6781} shows the final model image for
[NII]\,658.4\,nm plotted in contours over the observed image by
\cite{MPPS01}.  Again the good agreement of the apparent size, obtained
from the size of the fitted model grid and our determined distance, is
evident.

\begin{figure}[ht]
\includegraphics[width=\columnwidth]{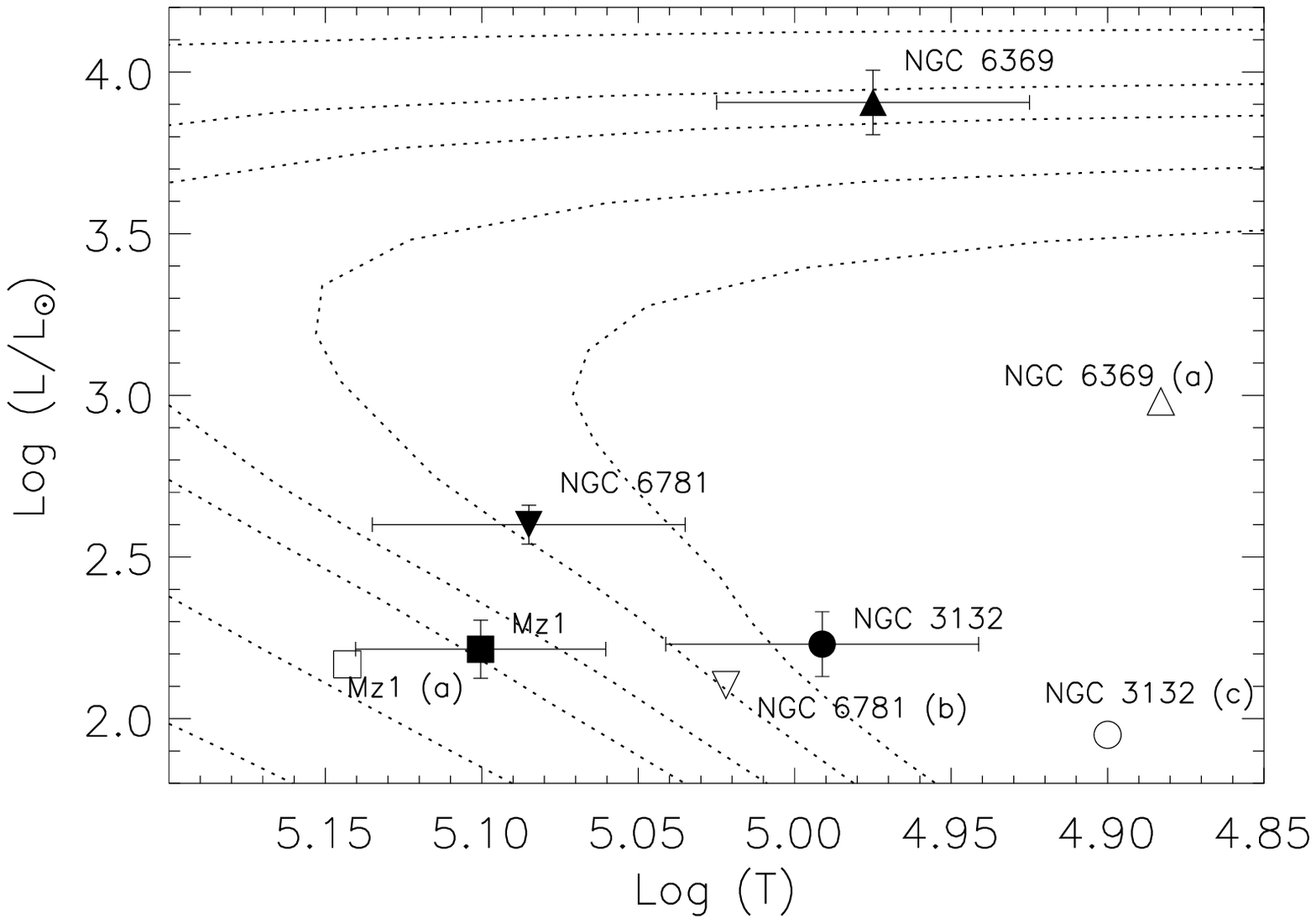}
\caption{HR diagram for NGC\,6781, NGC\,3132 (\cite{MMGV00}), 
NGC\,6369 (\cite{MSGH04}), MZ\,1 (\cite{MSGGH05}), all PNe that 
had their central star properties determined by our method. Also 
plotted are the literature values for comparison. a) \cite{SCS93}; b)
\cite{SVMG02}; c) . The evolutionary tracks are from \cite{VW94}; they 
are similar to the \cite{B95} models but take metalicities etc. into account.
. \label{evolallpn} }
\end{figure}

As with objects studied in previous works, the mass of the ionizing
star, as well as its progenitor and age are determined from
theoretical cooling tracks. Here we have used the cooling tracks of
\cite{VW94} because their grids present a good (=\,close) sampling of
progenitor masses in the 1 to 3$M_{\odot}$ range. In Figure
\ref{evolallpn} we show the position of NGC\,6781, as well as the
other objects that we have studied previously with this method, along
with the theoretical cooling tracks. From this we obtain the mass of
the central star of and its progenitor to be
$0.60\,\pm0.03\,M_{\odot}$ and $1.5\,\pm0.5\,M_{\odot}$
respectively. Figure\,\ref{evolallpn} also shows the position of these
same objects as determined by different authors by distinct
techniques. All PNe central stars are well evolved on their cooling
track except NGC\,6369.

\section{Discussion and conclusions}

As with all our other PNe central star temperatures determined by this
3-D structure method we generally find higher temperatures than those
computed from oxygen lines and nearer those from HeII
lines. Table\,\ref{param-comp} shows a comparison of parameters from
the literature and those determined by us in this and previous papers.

\begin{table*}[ht]
\begin{center}
\caption{Comparison of parameters found in the literature and those 
determined by our 3-D method. \label{param-comp}}
\begin{tabular}{lccccc}\hline {Object}  &   {Our T} & {HeII/HI} & 
{[OIII]/[OII]} & {Our L(L$_{\odot}$)} & {Our d(pc)}\\
\hline\hline 
\\
{Hb 5    } & 230$^a$ & 131$^b$ & -  &  6000$^a$ & 1400$^a$ \\
{Mz1     } & 120$^c$ & 139$^d$ & -  &   164$^c$ & 1050$^c$ \\
{NGC 3132} &  90$^e$ &  80$^c$ & 36$^c$ &   150$^e$ &  930$^e$ \\
{NGC 6369} &  91$^f$ & 122$^g$ & 60$^g$ &  8100$^f$ & 1550$^f$ \\
{NGC 6781} & 123$^h$ & 126$^g$ & 68$^g$ &   385$^h$ &  950$^h$ \\

\hline

\end{tabular}
\end{center}
\small{
a) Preliminary results from \cite{RSM04}
b) \cite{KJ89}
c) \cite{Ph03}
d) \cite{MSGGH05}
e) \cite{MMGV00}
f) \cite{MSGH04}
g) \cite{G97}
h) This paper.}
\end{table*}

Note that \cite{Ph03} found that bipolar PNe have average
T(HeII)\,=\,138k; ellipticals 92\,k; round 81\,k whole sample
87\,k. This was also found earlier by \cite{CS95} who determined
T(bipolars)\,=\,142\,k; irregulars 99\,k; ellipticals 76\,k; and
unresolved 63\,k. This general trend is always observed, and our high
temperature for bipolar NGC\,6781 is no exception.

The central star properties of all PNe that we have applied our method
to are shown in the HR diagram of Figure\,\ref{evolallpn}. Also shown
are the values determined by other methods, taken from the
literature. There are large differences between our values and those
previously published for NGC\,6781 and NGC\,6369 and in both cases the
central star luminosity determined by our method was higher than the
literature value. This is because it was assumed that these nebulae
are radiation bound but we have shown them to be matter bound as they
lose up to 70\% of their UV radiation to space, resulting in an
underestimation of both the luminosity and temperature. The blue rise
in the SED for NGC\,6781 also confirms that blue radiation is escaping
from the nebula. Note that the central star luminosity from our model,
L\,=\,385\,$/L_{\odot}$ is larger than the luminosity derived from the
observed SED L\,=\,225\,$/L_{\odot}$ by a factor of 1.7, confirming
that the nebula is matter bound and that a significant fraction of
the stellar UV flux escapes from the object.

We claim that ours are the most accurate luminosities and temperatures
that have been determined for these stars to date. Interestingly our
values tend to bring the core masses closer to $0.6~M_{\odot}$ which
is also the peak value of the narrow mass distribution of white
dwarfs.

\begin{acknowledgements}
      We acknowledge the continuing support of NOAO's Science Fund to
      HM and the generous hospitality of the Nordic Optical Telescope
      on La Palma.
\end{acknowledgements}


\begin{thebibliography}

\bibitem[Acker et al. (1992)]{AO92} Acker, A., Ochsenbein, F., Stenholm, 
B., Tylenda, R., Marcout, J., Schohn, C. 1992, Strabourg-ESO Catalogue of 
Galactic PNe.

\bibitem[Aller (1987)]{A87} Aller, L.H. Physics of Thermal Gaseous Nebulae, Reidel.

\bibitem[Bl\"{o}cker (1995)]{B95} Bl\"{o}cker, T.\ 1995, \aap, 299, 755 

\bibitem[Cahn, Kaler \& Stanghelini (1992)]{CKS92} Cahn, J.H., Kaler, 
J.B.; \& Stanghellini, L. 1992, \aaps, 94, 399

\bibitem[Ciardullo (2003)]{C03} Ciardullo, R. 2003, Lecture Notes in Physics, 635, 243

\bibitem[Corradi et al. (2003)]{CSSP03} Corradi, R.L.M., Sch\"{o}nberner, 
D., Steffen, M., Perinotto, M. 2003, MNRAS, 340, 417

\bibitem[Corradi \& Schwarz (1995)]{CS95} Corradi, R.L.M., Schwarz, 
H.E. 1995, \aap, 293, 871

\bibitem[Ercolano et al. (2003)]{EBSL03} Ercolano, B., Barlow, 
M.~J., Storey, P.~J., \& Liu, X.-W.\ 2003, \mnras, 340, 1136 

\bibitem[Feldmeier et al. (2004)]{FCJD04} Feldmeier, J.J., Ciardullo, R., 
Jacoby, G.H., Durrell, P.R. 2004, \apj, 615, 196

\bibitem[Filippenko (1982)]{F82} Filippenko, A. V.\ 1982, 
\pasp, 94, 715 

\bibitem[Gruenwald, Viegas \& Brogui\`{e}re (1997)]{GVB97} 
Gruenwald, R., Viegas, S.~M., \& Brogui\`{e}re, D.\ 1997, \apj, 480, 283 

\bibitem[Gurzadyan (1997)]{G97} Gurzadyan, G.A. 1997 ``The Physics and 
Dynamics of Planetary Nebulae'', Springer

\bibitem[Huggins, Bachiller, Cox \& Forveille (1996)]{HBCF96} Huggins, 
P.J., Bachiller, R., Cox, P., Forveille, T. 1996, \aap, 315, 284 

\bibitem[Iben \& Renzini (1983)]{IR83} Iben, I., \& Renzini, 
A.\ 1983, \araa, 21, 271

\bibitem[Jacoby et al. (1999)]{JCF99} Jacoby, G.~H., 
Ciardullo, R., \& Feldmeier, J.~J.\ 1999, ASP Conf.~Ser.~167: Harmonizing 
Cosmic Distance Scales in a Post-HIPPARCOS Era, 167, 175

\bibitem[Kaler \& Jacoby (1989)]{KJ89} Kaler, J.B., Jacoby, G.H. 
1989 AJ 345, 871

\bibitem[Kaler, Shaw \& Browning (1997)]{KSB97} Kaler, J.B., Shaw, R.A., 
Browning, L. 1997, \pasp, 109, 289

\bibitem[Liu, Liu, Luo \& Barlow (2004)]{LLLB04} Liu, Y., Liu, X.-W., 
Luo, S.-G., Barlow, M. J. 2004, MNRAS 353, 1231 

\bibitem[Maciel \& Costa (2003)]{MC03} Maciel, W.J., Costa, R.D.D. 
2003 IAUS 209, 551 Eds. Kwok et al.

\bibitem[Marston et al. (1998)]{MB98} Marston, A.P., 
Bryce, M., L\'opez, J.A., Palmer, J.W., Meaburn, J. 1998, \aap, 329,
683

\bibitem[Mavromatakis et al. (2001)]{MPPS01} Mavromatakis, F., 
Papamastorakis, J., Paleogolou, E.V. 2001, \aap, 280, 287

\bibitem[McCall (1984)]{M84}McCall, M. L. 1984, MNRAS, 208, 253

\bibitem[Monteiro, Morisset, Gruenwald \& Viegas (2000)]{MMGV00} Monteiro, 
H., Morisset, C., Gruenwald, R., \& Viegas, S.~M.\ 2000, \apj, 537, 853

\bibitem[Monteiro, Gruenwald, Morisset \& Viegas (2002)]{MMGV02}
Monteiro, H., Gruenwald, R., Morisset, C., \& Viegas, S.~M.\ 2002,
Revista Mexicana de Astronomia y Astrofisica Conference Series, 12, 170

\bibitem[Monteiro, Schwarz, Gruenwald \& Heathcote (2004)]{MSGH04} 
Monteiro, H., Schwarz, H.E., Gruenwald, R., \& Heathcote,S.R. 2004, 
\apj, 609,194

\bibitem[Monteiro, Schwarz, Gruenwald, Guenthner \& 
Heathcote (2005)]{MSGGH05} Monteiro, H., Schwarz, H.E., Gruenwald, R., 
Guenthner, K., \& Heathcote,S.R. 2005, \apj, 620,321

\bibitem[Myers et al. (1987)]{MY87} Myers, P. C.; Fuller, Gary A.; 
Mathieu, R. D.; Beichman, C. A.; Benson, P. J.; Schild, R. E.; 
Emerson, J. P. 1987, \apj, 319, 340

\bibitem[Osterbrock (1989)]{OS89} Osterbrock, D. E.\ 1989, 
Research supported by the University of California, John Simon 
Guggenheim Memorial Foundation, University of Minnesota, et al.~Mill 
Valley, CA, University Science Books  

\bibitem[Perek (1971)]{P71} Perek, L. 1971, Bull.Astr.Inst.Cz., 22, 103

\bibitem[Phillips (2003)]{Ph03} Phillips, J.P. 2003, MNRAS 344, 501

\bibitem[Pottasch (1984)]{P84} Pottasch, S.R. 1984, Planetary Nebulae, Reidel

\bibitem[Rauch, Deetjen, Dreizler \&  Werner (2000)]{RDDW00} Rauch, T., 
Deetjen, J. L., Dreizler, S. e Werner, K. 2000, Asymmetrical Planetary 
Nebulae II: From origins to microstructures, ASP Conference Series 199, 337 

\bibitem[Rice, Schwarz \& Monteiro (2004)]{RSM04} Rice, M., 
Schwarz, H.E., Monteiro, H. 2004 AAS Abs. 20513813

\bibitem[Schwarz, Corradi \& Melnick (1992)]{SCM92}Schwarz, H. E., 
Corradi, R. L. M., \& Melnick, J. 1992, AApS, 96, 23
 
\bibitem[Schwarz, Aspin, Corradi, Reipurth \& (1997)]{S97}Schwarz, H. 
E., Aspin, C., Corradi, R. L. M., Reipurth, B.  1997, \aap, 319, 267
 
\bibitem[Seaton (1979)]{S79}Seaton, M. J. 1979, MNRAS, 187, 73 

\bibitem[Stanghellini, Corradi \& Schwarz (1993)]{SCS93} Stanghellini, L., 
Corradi, R.L.M., Schwarz, H.E. 1993, \aap, 279,521

\bibitem[Stanghellini, Villaver, Manchado \& Guerrero (2002)]{SVMG02}
Stanghellini, L., Villaver, E., Manchado, A., \& Guerrero, M. A. 
2002, \apj, 576, 285

\bibitem[Van de Steene \& Zijlstra (1995)]{VZ95} Van de Steene, G.C., 
Zijlstra, A.A., 1995, \aap, 293, 541

\bibitem[Vassiliadis \& Wood (1994)]{VW94}Vassiliadis, E., Wood, P.R. 
1994, \apjs, 92, 125 

\bibitem[Webster, Payne, Storey \& Dopita (1988)]{W88} Webster, B. L.; 
Payne, P. W.; Storey, J. W. V.; Dopita, M. A. 1988, MNRAS, 235, 533

\bibitem[Zhang (1995)]{Z95} Zhang, C. Y. 1995, \apjs, 98, 659

\end{thebibliography}
\end{document}